\documentclass{article} 

\usepackage[english]{babel} 
\usepackage{amssymb}
\usepackage{amsmath}
\usepackage{txfonts}
\usepackage{mathdots}
\usepackage[classicReIm]{kpfonts}
\usepackage[pdftex]{graphicx}
\usepackage{geometry} 
\begin{document}

\title{%
  Itemsets of interest for negative association rules 
  }

\author{Hyeok Kong, Dokjun An, Jihyang Ri \\
	Faculty of Mathematics, Kim Il Sung University, Pyongyang, D. P. R. K. \\
	hyeok\_kong@yahoo.com\\
	matadj79@yahoo.com\\
	rjh0415@yahoo.com
	}

\maketitle

\noindent \textbf{Abstract:} So far, most of association rule minings have considered about positive association rules based on frequent itemsets in databases\cite{Kong2015,Piatetsky1991,Agrawal1994,Webb2000}, but they have not considered the problem of mining negative association rules correlated with frequent and infrequent itemsets. Negative association rule mining is much more difficult than positive association rule mining because it needs infrequent itemsets, and only the rare association rule mining which is a kind of negative association rule minings has been studied. 

\noindent This paper presents a mathematical model to mine positive and negative association rules precisely, for which in a point of view that negation of a frequent itemset is an infrequent itemset, we make clear the importance of the problem of mining negative association rules based on certain infrequent itemsets and study on what conditions infrequent itemsets of interest should satisfy for negative association rules.


\section{Related works}

\noindent Apriori, a traditional association rule mining algorithm, was first proposed in 1993\cite{Agrawal1994}.The Apriori algorithm scans database multiple times, extracts a large number of possible frequent itemsets, and generates association rules. This repeated candidate generation-test approach can impose large computational overheads on a computer when the number of frequent itemsets is very large. To overcome this difficulty, a novel frequent pattern mining model based on an FP-tree was proposed in 2000\cite{Webb2000}. FP-growth algorithm based on the FP-tree was known by many people later and was widely used. And an OPUS-search was also proposed in 2000\cite{Webb2000}, which reduces the search space on basis of interrelationships among itemsets. Experiments showed that the model based on FP-tree and the OPUS-search were more efficient than the Apriori algorithm and its derivatives.

\noindent By the way, some itemsets pruned in the algorithms of mining frequent patterns might be useful for mining negative association rules. That is, in spite of the fact that frequent itemsets are not contained in any association rules, they should not be pruned. These frequent itemsets might be as helpful as ever to mining negative association rules. So the existing pruning techniques have not come true to mine negative association rules.

\noindent Recently, many researches have been done in the area of rare pattern mining\cite{Kong2016,Almasi2015}. Principally, there are two types of researches such as level-wise and tree-based, like that of current frequent pattern mining. More recently, a new approach of list structure was proposed\cite{Jain2015}. MS-Apriori, Rarity, ARIMA, AfRIM and Apriori-Inverse are 5 algorithms of mining rare itemsets\cite{Almasi2015}.

\noindent Note that rare association rule such as $\lnot$A$\mathrm{\to}$$\mathrm{\neg}$C is only a kind of negative association rules, meanwhile any clear definition of all possible negative association rules have not given. 

\noindent And, there are much more infrequent itemsets for negative association rules than frequent itemsets(almost double) and the size of the search space may be exponential\cite{Almasi2015,Jain2015}. It has not been studied what conditions infrequent itemsets satisfy could be useful to generate negative association rules.

\section{The concept of negative association rules.}

\noindent In decision making, we generally use many advantage factors, and also consider some disadvantage foctors for the purpose of low-risk(high-profit). Specially, it is important to capture which of disadvantage factors rarely occur when the expected advantage facors occur in using past data. Similarly, it is necessary to focus on not only positive association rule mining, but also negative association rule mining, but there are essential differences between them.

\noindent Traditional association rule minings mainly identify strong rules among frequent and high correlated itemsets, and the extracted association rules are \textbf{positive rules}.

\noindent In applications, an association rule A$\mathrm{\to}$B is used to predict that "if A occurs, then B also generally occurs." This association rule is applied to place "B near A" for efficiency in applications such as store layout, product placement and supermarket management. For store layout, associations analysis suggests that B should be placed near to A if there is a rule A$\mathrm{\to}$B for itemsets A and B. 

\noindent In contrast, the rules of the form A$\mathrm{\to}$$\mathrm{\neg}$B(or $\mathrm{\neg}$A$\mathrm{\to}$B, or $\mathrm{\neg}$A$\mathrm{\to}$$\mathrm{\neg}$B) are called \textbf{negative rules}. Negative rules indicate that the presence of some itemsets will imply the absence of other itemsets.

\noindent To focus on not only positive association rule mining but also negative association rule mining is very important because both of advantage and disadvantage factors should be considered in decision making for applications including product placement and investment analysis. Negative association rules such as A$\mathrm{\to}$$\mathrm{\neg}$B mean that B(disadvantage factor) rarely occurs when A(advantage factor) occurs, and these rules are useful in decision making. 

\noindent And, to focus on not only positive association rule mining but also negative association rule mining is also very important because negative relationships play an important role in several areas of science and technology (as negative numbers in mathematics and negation in logic). The association rule such as A$\mathrm{\to}$$\mathrm{\neg}$B describes another type of relationship among itemsets: negation. Therefore, negative association rules can be as important as positive rules in association analysis.

\noindent From the above reasons, we study to find negative association rules too when identifying positive association rules in databases. The following example shows us how to identify negative association rules hidden in databases.

\noindent \textbf{Example. }The purchase data of soy and salt in a supermarket are showed in Table 1. For an association rule "soy$\mathrm{\to}$salt", suppose sprt(soy)=0.25 and sprt(soy$\mathrm{\cup}$salt)=0.2.\textbf{}

Table 1. The purchase data of soy and salt

\begin{tabular}{|p{0.9in}|p{0.6in}|p{0.5in}|p{0.5in}|} \hline 
 & salt & $\mathrm{\neg}$salt & $\mathrm{\sum}$row \\ \hline 
soy & 20 & 5 & 25 \\ \hline 
$\mathrm{\neg}$soy & 70 & 5 & 75 \\ \hline 
$\mathrm{\sum}$col & 90 & 10 & 100 \\ \hline 
\end{tabular}

Now, we apply the support-confidence model to a potential association rule "soy$\mathrm{\to}$salt", then the support of the rule is fairly high as 0.2. The confidence is the conditional probability that a customer buys salt, given that he/she buys soy, i.e. sprt(soy$\mathrm{\cup}$salt)/sprt(soy)=0.2/0.25=0.8, which is also fairly high. At this point, we may conclude that the rule "soy$\mathrm{\to}$salt" is a valid rule.

\noindent In Table 2, we suppose sprt(salt)=0.6, sprt(soy)=0.4, sprt(soy$\mathrm{\cup}$salt)=0.05 and minconf=0.52. Then, the conditional probability is as follows.

\noindent sprt(soy$\mathrm{\cup}$salt)/sprt(soy)=0.05/0.4$<$minconf=0.52

Table 2. The updated purchase data of soy and salt

\begin{tabular}{|p{0.9in}|p{0.6in}|p{0.5in}|p{0.5in}|} \hline 
 & salt & $\mathrm{\neg}$salt & $\mathrm{\sum}$row \\ \hline 
soy & 5 & 35 & 40 \\ \hline 
$\mathrm{\neg}$soy & 55 & 5 & 60 \\ \hline 
$\mathrm{\sum}$col & 60 & 40 & 100 \\ \hline 
\end{tabular}

And, the support of soy$\mathrm{\cup}$salt is as low as sprt(soy$\mathrm{\cup}$salt)=0.05. This means that   soy$\mathrm{\cup}$salt is an infrequent itemset and "soy$\mathrm{\to}$salt" cannot be extracted as a rule in support-confidence model.

\noindent However, "soy$\mathrm{\to}$$\mathrm{\neg}$salt" can be extracted as a negative rule in the database from the following expressions.

\noindent sprt(soy$\mathrm{\cup}$$\mathrm{\neg}$salt) = sprt(soy) - sprt(soy$\mathrm{\cup}$salt) = 0.4 - 0.05 = 0.35

\noindent sprt(soy$\mathrm{\cup}$$\mathrm{\neg}$ salt)/sprt(soy) = 0.35/0.4 = 0.875 $>$ minconf

\noindent From the above example, we need to search for infrequent itemsets to mine negative association rules.

\noindent Therefore, \textbf{infrequent itemsets} in databases should be examined to mine negative association rules. And, as you know, the previous algorithms are generally involved with identifying only\textbf{ frequent itemsets} in a given database, so that they are inadequate for mining of negative association rules.

\noindent On the other hand, the search space of infrequent itemsets has an exponential size because items are randomly combined in transactions of the database, and the search space problem arises in identifying infrequent itemsets of interest. For this reason, it is needed to study a new method to choose not only \textbf{positive itemsets of interest} but also \textbf{negative ones}, to reduce the size of the search space.

\section{Itemsets of interest}

\noindent As described above, there are so many infrequent itemsets of almost exponential amount in databases and only some of them are useful to negative association rules of interest. So, it should be made clear what negative association rules are useful to applications.

\noindent 

\noindent \textbf{1)} \textbf{Positive itemsets of interest}

\noindent 

\noindent Previous works on the amount of `usefulness' or `interest' of a rule focused on how much the actual support of a rule exceeded the expected support, based on the support of the antecedent and consequent. When the support of X is denoted by sprt(X), Piatetsky-Shapiro${}^{[5]}$${}^{ }$argued that a rule X$\mathrm{\to}$Y is not interesting if the following expression is satisfied.

\noindent sprt(X$\mathrm{\to}$Y)$\mathrm{\approx}$sprt(X)$\times $sprt(Y)

\noindent This argument is significant in probabilistic level, because it is a statistical definition of dependence for the itemsets X and Y, so we take this expression as a condition for judging the itemsets of interest. 

\noindent When the confidence of X$\mathrm{\to}$Y is denoted by conf(X$\mathrm{\to}$Y), conf(X$\mathrm{\to}$Y) and sprt(X$\mathrm{\cup}$Y) can be represented based on probabilistic theory as follows.

\noindent sprt (X$\mathrm{\cup}$Y)=p(X$\mathrm{\cup}$Y)

\noindent conf(X$\mathrm{\to}$Y)=p(Y{\textbar}X)= $\frac{p(X\bigcup Y)}{p(X)} $

\noindent Then, the argument of Piatetsky-Shapiro can be represented as follows.

\noindent p(X$\mathrm{\cup}$Y)$\mathrm{\approx}$p(X)p(Y)

\noindent Based on this expression, we define the measure of interest of a rule, denoted by Interest(X, Y), as in [1]. It is 

\noindent Interest(X, Y) = sprt(X$\mathrm{\cup}$Y)/(sprt(X)sprt(Y)).

\noindent That is, if Interest(X, Y) =1, then X and Y are independent, so the rule X$\mathrm{\to}$Y is not interesting. This means that if the antecedent and the consequent are independent, then the rule is not interesting. Clearly, the further the value of Interest(X, Y) is away from 1, the more the dependence of X and Y is.

\noindent \textbf{Theorem1.} (The argument of Piatetsky-Shapiro) Let I be a set of all items in a database, X,Y$\mathrm{\subseteq}$I be itemsets, X$\mathrm{\cap}$Y=$\mathrm{\emptyset }$, sprt(X)$\mathrm{\neq}$0, and sprt(Y)$\mathrm{\neq}$0. Also, the thresholds: minsprt(minimal support), minconf(minimal confidence) and mininterest(minimal interest) $>$ 0 are given by users or experts. Then, the rule X$\mathrm{\to}$Y can be extracted as \textbf{an association rule}\textit{ }if

\begin{enumerate}
\item  sprt(X$\mathrm{\cup}$Y) $\mathrm{\ge}$ minsprt,  

\item  sprt(Y{\textbar}X) $\mathrm{\ge}$ minconf,

\item  {\textbar}sprt(X$\mathrm{\cup}$Y) $-$ sprt(X)sprt(Y){\textbar} $\mathrm{\ge}$ mininterest.
\end{enumerate}

\noindent Proof. (3) can be represented as follows.
\[\frac{{\rm |}sprt(X\bigcup Y){\rm -}sprt(X)sprt(Y){\rm |}}{sprt(X)sprt(Y)} |{\rm \; }\ge \frac{\min interest}{sprt(X)sprt(Y)}  ,\] 
\[{\rm |}\frac{sprt(X\bigcup Y)}{sprt(X)sprt(Y)} {\rm -1}|{\rm \; }\ge \frac{\min interest}{sprt(X)sprt(Y)} \] 
Since $sprt(X)sprt(Y)\le {\rm 1}$, we have 
\[\frac{\min interest}{sprt(X)sprt(Y)} {\rm \; }\ge {\rm \; }\frac{\min interest}{{\rm 1}} =\min interest.\] 
Hence,
\[|\frac{sprt(X\bigcup Y)}{sprt(X)sprt(Y)} -1|\ge \min interest.\] 
By the definition of interest, X$\mathrm{\to}$Y can be extracted as a rule of interest. (End)

\noindent There could be different values of minimum interests in different applications. In the above example, mininterest=0.08 is a valid value, but when all supports of the itemsets are less than 0.08, this is invalid. Therefore, a proper minimum interest should be chosen for an application by users or experts. If sprt(X$\mathrm{\cup}$Y)=minsprt for the itemset X$\mathrm{\cup}$Y, then X$\mathrm{\to}$Y could be possibly extracted as a rule of interest. Although  sprt(X$\mathrm{\cup}$Y)=sprt(X)=sprt(Y)=minsprt, but X$\mathrm{\to}$Y could be also extracted as a rule of interest. From these facts, the following condition is satisfied for minimum interest.

\noindent {\textbar}sprt(X$\mathrm{\cup}$Y)$-$sprt(X)sprt(Y){\textbar} = minsprt - minsprt${}^{2}$ $\mathrm{\ge}$ mininterest.

\noindent Therefore, minsprt-minsprt${}^{2}$ can be taken as an upper bound of mininterest. For example, if minsprt=0.2 and minsprt=0.001, then the following expressions are respectively satisfied.

\noindent mininterest $\le $ minsprt - minsprt${}^{2}$ = 0.2 -0.2${}^{2}$${}^{ }$= 0.16,

\noindent mininterest $\le $ minsprt - minsprt${}^{2}$ = 0.001 -0.0012${}^{2}$${}^{ }$= 0.00099.

\noindent Theorem 1 shows that if sprt(X$\mathrm{\cup}$Y)$\mathrm{\approx}$sprt(X)sprt(Y), then X$\mathrm{\to}$Y cannot be extracted as a rule of interest. In fact, sprt(X$\mathrm{\cup}$Y)$\mathrm{\approx}$sprt(X)sprt(Y) means that X and Y are almost independent in probability theory terms. Generally, if sprt(X$\mathrm{\cup}$Y)$-$sprt(X)sprt(Y) $\mathrm{\ge}$ mininterest, then the rule X$\mathrm{\to}$Y is of interest. Thus, we can give the following definition.

\noindent \textbf{Definition 1.} X$\mathrm{\to}$Y is called \textbf{a positive association rule of interest}, and X$\mathrm{\cup}$Y is called \textbf{a positive itemset of interest}, if they satisfy the following conditions.

\begin{enumerate}
\item  $X\bigcap Y=\phi $

\item  sprt(X$\mathrm{\cup}$Y) $\mathrm{\ge}$ minsprt ,  

\item  {\textbar}sprt(X$\mathrm{\cup}$Y) $-$ sprt(X)sprt(Y){\textbar} $\mathrm{\ge}$ mininterest,

\item  sprt(X$\mathrm{\cup}$Y)/sprt(X) $\mathrm{\ge}$ minconf
\end{enumerate}

\noindent Otherwise, if {\textbar}sprt(X$\mathrm{\cup}$Y)$-$sprt(X)sprt(Y){\textbar} $<$ mininterest or sprt(X$\mathrm{\cup}$Y)/sprt(X) $<$ minconf, then the rule X$\mathrm{\to}$Y is not of interest, and X$\mathrm{\cup}$Y is called \textbf{an uninteresting itemset}.

\noindent Conversely, if Q is a positive itemset of interest, there is at least one expression Q= X$\mathrm{\cup}$Y such that X and Y satisfy the above 4 conditions for positive association rules of interest.

\noindent 

\noindent \textbf{2)} \textbf{Negative itemsets of interest}

\noindent 

\noindent To mine negative association rules, all itemsets for such rules in a given database must be generated. For example, if A$\mathrm{\to}$$\mathrm{\neg}$B (or $\mathrm{\neg}$A$\mathrm{\to}$B, or $\mathrm{\neg}$A$\mathrm{\to}$$\mathrm{\neg}$B) can be found, then sprt(A$\mathrm{\cup}$$\mathrm{\neg}$B) $\mathrm{\ge}$ minsprt (or sprt($\mathrm{\neg}$A$\mathrm{\cup}$B) $\mathrm{\ge}$ minsprt, or sprt($\mathrm{\neg}$A$\mathrm{\cup}$$\mathrm{\neg}$B) $\mathrm{\ge}$ minsprt) must be satisfied. This means that sprt(A$\mathrm{\cup}$B) $<$ minsprt may be satisfied. And the itemset A$\mathrm{\cup}$B may not be generated as a frequent itemset when using conventional algorithms. That is, A$\mathrm{\cup}$B may be an infrequent itemset. However, the numbers of infrequent itemsets are usually so many that we must identify only the infrequent itemsets useful to applications. It is a key which infrequent itemsets are of interest. Therefore, we must define some conditions for identifying all infrequent itemsets of interest.

\noindent Generally, in a large scale database, if A is a frequent itemset and B is an infrequent itemset with frequency 1, then A$\mathrm{\to}$$\mathrm{\neg}$B is a valid negative rule. In fact, supp(A) $\mathrm{\ge}$ minsprt and sprt(B) $\mathrm{\approx}$ 0. So

\noindent sprt(A$\mathrm{\cup}$$\mathrm{\neg}$B)$\mathrm{\approx}$sprt(A) $\mathrm{\ge}$ minsprt

\noindent conf(A$\mathrm{\to}$$\mathrm{\neg}$B) = sprt(A$\mathrm{\cup}$$\mathrm{\neg}$B)/sprt(A)$\mathrm{\approx}$1 $\mathrm{\ge}$ minconf.

\noindent This means that the rule A$\mathrm{\to}$$\mathrm{\neg}$B is valid. There can be a number of itemsets of this kind in databases. For example, some rarely purchased goods in a store fit into this category. 

\noindent However, they are frequent itemsets that usually attract attention in applications. Hence, any patterns mined in databases would commonly relate to frequent itemsets. This means that if A$\mathrm{\to}$$\mathrm{\neg}$B (or $\mathrm{\neg}$A$\mathrm{\to}$B, or $\mathrm{\neg}$A$\mathrm{\to}$$\mathrm{\neg}$B) is a negative rule of interest, A and B would involve only frequent itemsets. This is one of the main conditions for identifying interesting negative association rules. 

\noindent \textbf{Definition 2.} If there is at least one expression Q= A$\mathrm{\cup}$B such that A and B satisfy the following 3 conditions for negtive association rule of type A$\mathrm{\to}$$\mathrm{\neg}$B (or $\mathrm{\neg}$A$\mathrm{\to}$B, or $\mathrm{\neg}$A$\mathrm{\to}$$\mathrm{\neg}$B), then the itemset Q is called \textbf{a negative itemset}.

\begin{enumerate}
\item  $A\bigcap B=\phi ,$

\item  sprt(A) $\mathrm{\ge}$ minsprt , sprt(B) $\mathrm{\ge}$ minsprt ,

\item  sprt(A$\mathrm{\cup}$$\mathrm{\neg}$B) $\mathrm{\ge}$ minsprt (or sprt($\mathrm{\neg}$A$\mathrm{\cup}$B) $\mathrm{\ge}$ minsprt, or sprt($\mathrm{\neg}$A$\mathrm{\cup}$$\mathrm{\neg}$B) $\mathrm{\ge}$ minsprt). 
\end{enumerate}

\noindent The probability significance of negative association rules can be guaranteed by the above condition (2). The other conditions guarantee the rules valid.

\noindent For the Piatetsky-Shapiro argument, the rule A$\mathrm{\to}$$\mathrm{\neg}$B is of interest if sprt(A$\mathrm{\cup}$$\mathrm{\neg}$B) -- sprt(A)sprt($\mathrm{\neg}$B) $\mathrm{\ge}$ mininterest. Therefore, we can give the following definition.

\noindent \textbf{Definition 3.}  The rule A$\mathrm{\to}$$\mathrm{\neg}$B is called \textbf{a negative association rule of interest} and A$\mathrm{\cup}$B is called \textbf{a negative itemset of interest} if they satisfy the following conditions.

\begin{enumerate}
\item  $A\bigcap B=\phi ,$

\item  sprt(A) $\mathrm{\ge}$ minsprt , sprt(B) $\mathrm{\ge}$ minsprt , sprt(A$\mathrm{\cup}$$\mathrm{\neg}$B) $\mathrm{\ge}$ minsprt,

\item  sprt(A$\mathrm{\cup}$$\mathrm{\neg}$B) $-$ sprt(A)sprt($\mathrm{\neg}$B) $\mathrm{\ge}$ mininterest,

\item  sprt(A$\mathrm{\cup}$$\mathrm{\neg}$B)/sprt(A) $\mathrm{\ge}$ minconf.
\end{enumerate}

\noindent Otherwise, the rule A$\mathrm{\to}$$\mathrm{\neg}$B is not of interest, and A$\mathrm{\cup}$B is \textbf{an uninteresting itemset}. Thus, uninteresting itemsets are any itemsets in a database which exclude both positive and negative itemsets of interest. These itemsets need to be pruned to reduce the space searched in mining.

\noindent On the other hand, if Q is a negative itemset of interest, there is at least one expression Q=A$\mathrm{\cup}$B such that one of the rules: A$\mathrm{\to}$$\mathrm{\neg}$B, or $\mathrm{\neg}$A$\mathrm{\to}$B, or $\mathrm{\neg}$A$\mathrm{\to}$$\mathrm{\neg}$B, is a valid negative association rule of interest.

\noindent In a few words, there are many frequent itemsets related to uninteresting association rules. Therefore, if we extract only positive and negative itemsets of interest, the search space can be extremely reduced.

\section{Conclusions}

\noindent This paper presented a mathematical model to mine positive and negative association rules precisely, for which in such a point of view that negation of a frequent itemset is an infrequent itemset, we made clear the importance of the problem of mining negative association rules based on certain infrequent itemsets, studied on what conditions infrequent itemsets of interest should satisfy for negative association rules and defined positive and negative itemsets of interest, to reduce the size of the search space.

\noindent The mining of negative association rules is reasonable from two aspects. First, the complete relationships among itemsets are considered, to establish a scientific system  of positive and negative rules just as negative real number system introduced in applications with positive real number system. Second, it can provide more information for decision making in applications. 

\noindent For a further work, it needs to develop algorithms to search only for positive and negative itemsets of interest in a given database, based on the mathematical model presented in this paper.

\noindent \textbf{}

\noindent \textbf{References}

\bibliographystyle{plain}
\bibliography{reference}

\begin{thebibliography}{1}

\bibitem{Kong2015}
C.~Jong. H.~Kong.
\newblock Improving efficiency of an algorithm for discovering frequent
  itemsets.
\newblock {\em International Journal of Theoretical Physics and Cryptography},
  10(12):1--5, 2015.

\bibitem{Kong2016}
U.~Ryang. H.~Kong, C.~Jong.
\newblock Rare association mining for network intrusion detection.
\newblock {\em International Journal of Theoretical Physics and Cryptography},
  12(12):13--17, 2016.

\bibitem{Almasi2015}
Rare-PEARs. M.~Almasi, M. S.~Abadeh.
\newblock A new multi-objective evolutionary algorithm to mine rare and
  non-redundant quantitative association rules.
\newblock {\em Knowledge-Based Systems}, 89:366--384, 2015.

\bibitem{Piatetsky1991}
G.~Piatetsky-Shapiro.
\newblock Discovery, analysis, and presentation of strong rules.
\newblock {\em Knowledge Discovery in Databases, G. Piatetsky-Shapiro and W.
  Frawley (Eds.), AAAI Press/MIT Press}, pages 229--248, 1991.

\bibitem{Agrawal1994}
R.~Srikant. R.~Agrawal.
\newblock Fast algorithms for mining association rules.
\newblock {\em Proceedings of International Conference on Very Large Data
  Bases}, pages 487--499, 1994.

\bibitem{Jain2015}
R.~G.~Vishwakarma. S.~Jain.
\newblock Generating positive \& negative rules using apriori algorithm.
\newblock {\em Proceedings of Third IRF International Conference on 8th
  February 2015}, pages 34--38, 2015.

\bibitem{Webb2000}
G.~Webb.
\newblock Efficient search for association rules.
\newblock {\em Proceedings of International Conference on Knowledge Discovery
  and Data Mining}, pages 99--107, 2000.

\end{thebibliography}

\end{document}